# Data-Driven Power Flow Linearization: A Regression Approach

Yuxiao Liu, *Student Member, IEEE,* Ning Zhang, *Member, IEEE,* Yi Wang, *Student Member, IEEE,* Jingwei Yang, *Student Member, IEEE,* and Chongqing Kang, *Fellow, IEEE*

*Abstract*—The linearization of a power flow (PF) model is an important approach for simplifying and accelerating the calculation of a power system's control, operation, and optimization. Traditional model-based methods derive linearized PF models by making approximations in the analytical PF model according to the physical characteristics of the power system. Today, more measurements of the power system are available and thus facilitate data-driven approaches beyond model-driven approaches. This work studies a linearized PF model through a data-driven approach. Both a forward regression model (($P,Q$) as a function of ($\theta,V$)) and an inverse regression model (($\theta,V$) as a function of ($P,Q$)) are proposed. Partial least square (PLS)- and Bayesian linear regression (BLR)-based algorithms are designed to address data collinearity and avoid overfitting. The proposed approach is tested on a series of IEEE standard cases, which include both meshed transmission grids and radial distribution grids, with both Monte Carlo simulated data and public testing data. The results show that the proposed approach can realize a higher calculation accuracy than model-based approaches can. The results also demonstrate that the obtained regression parameter matrices of data-driven models reflect power system physics by demonstrating similar patterns with some power system matrices (e.g., the admittance matrix).

## I. INTRODUCTION

Power flow (PF) analysis is the basis of power system analysis and optimization. The nonlinearity of PF equations leads to difficulties in optimization and control algorithms [1], as in non-convergence problems, and incurs high computation burdens. The linearization of PF equations can markedly simplify the complexity and ensure the convergence of algorithm, which is why it is already widely used in power system control [2], scheduling [3] and market clearing [4, 5]. Among all of the PF linearization approaches, the DC power flow (DCPF) equations are currently the most widely used in industry. DCPF reveals the approximated linearity relationship between active power injection ($P$) and phase angle ($\theta$). A substantial number of studies has been conducted to enhance the DCPF and have considered the formulation of reactive power injections ($Q$) and voltage magnitudes ($V$) [6-8]. In [6], PF equations are formulated as a linearized form with

This work was supported in part by the National Key R&D Program of China (No. 2016YFB0900100), the National Science Foundation of China (No. 51620105007), and the Scientific & Technical Project of State Grid. (Corresponding author: Chongqing Kang.)
Y. Liu, N. Zhang, Y. Wang, J. Yang, and C. Kang are with the State Key Lab of Power Systems, Department of Electrical Engineering, Tsinghua University, Beijing 100084, China. (E-mail: cqkang@tsinghua.edu.cn).

respect to the square of the voltage magnitude ($V^2$) and modified phase angle ($V^2\theta$), which reveals a linearized PF relationship that can consider the reactive power. The model achieves acceptable accuracy, even under cold-start circumstances. On this basis, linearized models have been further proposed using the square of the voltage magnitude and phase angle [7] and the decoupled voltage magnitude and phase angle [8] as the independent variable. The above linearization methods improve the accuracy beyond DCPF.

With the spread of massive phasor measurement units (PMUs) and supervisory control and data acquisition (SCADA) systems, measurement data from power systems are sufficient to be used in rebuilding system models. The methods are known as data-driven methods and are found to contribute to the efficiency and accuracy of power system analysis. Traditionally in power system PF analysis, many model-based approaches based on a precise PF model are used to derive the models that facilitate rapid calculation or ensure optimization convergence. We consider these approaches to be model-to-model approaches. Different from model-to-model approaches, many data-driven methods rediscover model parameters from various operation data, which is denoted the data-to-model approach. Chen *et al* proposed a measurement-based method to estimate the distribution factors [9] and the Jacobian matrix [10] using the least square method. The rediscovered model parameters have advantages in near real-time flexibility to adapt to changes in topology or load. Yuan *et al.* identified the admittance matrix in a distribution network using graph theory [11]. The model can recover the real-time topology and admittance matrix in distribution grids with several hidden nodes without measurements. Few works have focused on the non-network-parameter-based data-driven method of PF calculations. To the best knowledge of the authors, the closest work is [12], which uses non-linear support vector regression (SVR) to reveal relationships among variables in a PF analysis. A nonlinear mapping rule between active and reactive bus injection ($P,Q$) and the phase angle and voltage magnitude ($V,\theta$) is built based on historical data. However, the phase angle and voltage magnitude are considered in a coupling form ( $\phi(V,\theta)$ ), which cannot consider different bus types in the mapping process. For example, the *PV* bus cannot be considered for the coupling of the phase angle and voltage magnitude.

Our work focuses on the data-driven linearization method for PF analysis. Compared with the current model-based PF linearization approaches mentioned above, the data-driven linearization PF analysis has the following advantages: 1) It does not require knowledge of the system topologies and

2parameters. In distribution grids, due to frequent re-configurations and increasing penetration of distributed energy resources, the exact system topologies, element parameters, and the control logic of active control devices are difficult to model accurately [12], [13]. Data-driven approaches are merely based on historical measurements and thus have significant advantages under these circumstances. 2) It improves the linearization accuracy of PF calculations. The training data reflects the real operation status of the power system such that the parameters of the data-driven approach more accurately consider the power system operation condition than model-based approaches do. For example, the data-driven approach can consider the deviation of parameters due to the atmospheric condition and aging [14].

Compared with the current data-driven approaches for PF analysis mentioned in [9]-[11], the proposed method has the following advantages: 1) Reducing calculation errors. The current data-driven approach identifies the parameters in the PF model first and then uses the identified model to conduct further control and optimizations. The PF calculation error may accumulate in the data-to-model and model-to-data processes. Our work replaces the process with a direct data-to-data approach and thus avoids modeling errors. 2) Reducing the computational burden. The current data-driven model still obtains a non-linearized PF model [12] and suffers from the computational burden for further applications, such as probabilistic load flow [15] and contingency analysis [16]. The proposed method reveals the linearized mapping relationship between operation variables based on historical data. 3) Enhancing computational flexibility. The proposed data-driven method can flexibly solve PF problems by considering different settings on bus types.

The contributions of this paper can be summarized as follows:

1) A data-driven linearization approach of PF equations is proposed that does not require knowledge of the power grid parameters and considers PF physics.

2) Both forward and inverse regression models of PF equations are produced that facilitates PF calculations with different settings of bus types.

3) Both partial least squares (PLS)- and Bayesian linear regression (BLR)-based algorithms are introduced to address the collinearity and avoid the overfitting of real operation data.

The remainder of this paper is organized as follows: Section II revisits the PF linearization problem from a data perspective and provides the framework of the data-driven PF linearization approach. Section III proposes forward and inverse models to regress linearized PF equations parameters. Section IV introduces PLS and BLR models. Section V validates the accuracy and robustness of the proposed model on several cases. Finally, conclusions are drawn in Section VI.

## II. Data-Driven Power Flow Linearization Framework

This section first interprets the PF calculation from a data-driven point of view. Then, the idea of the data-driven PF linearization is presented with discussions on its feasibility and challenges. The framework of data-driven PF linearization is finally proposed.

### A. PF Calculation from a Data-Driven Perspective

The steady state of a power system can be uniquely described by the power injections, node voltage, and branch power flow. The measurements of these quantities at a certain time period (e.g., at time $t$) can be formulated as a vector $\mathbf{x}_t$. The expression of $\mathbf{x}_t$ is shown in Equation (1), where $P_i^t$, $Q_i^t$, $V_i^t$ and $\theta_i^t$ represent the active power injection, reactive power injection, voltage magnitude and voltage angle of the bus $i$ at time $t$, respectively; $PF_i^t$ and $QF_i^t$ represent the active and reactive power flow of branch $l$ at time $t$, respectively.

$$\mathbf{x}_t = (P_{1,t}, \ ..., \ P_{N,t}, \ Q_{1,t}, \ ..., \ Q_{N,t}, \\ V_{1,t}, \ ..., \ V_{N,t}, \ \theta_{1,t}, \ ..., \ \theta_{N,t}, \quad (1) \\ PF_{1,t}, \ ..., \ PF_{L,t}, \ QF_{1,t}, \ ..., \ QF_{L,t}, \ ...)$$

Vector $\mathbf{x}_t$ represents a power system operation snapshot in time $t$. From the data point of view, it can be seen as a high-dimensional vector in a high-dimensional hyperspace. All of the vectors $\mathbf{x}_t$ that describe different operation states of a power system are on the high-dimensional surface described by the nonlinear AC power flow (ACPF) equations in Equation (2):

$$\begin{aligned} P_i &= V_i \sum_{j \in i} V_j (G_{ij} \cos \theta_{ij} + B_{ij} \sin \theta_{ij}) \\ Q_i &= V_i \sum_{j \in i} V_j (G_{ij} \sin \theta_{ij} - B_{ij} \cos \theta_{ij}) \\ PF_{ij} &= (V_i^2 - V_i V_j \cos \theta_{ij}) g_{ij} - V_i V_j \sin \theta_{ij} b_{ij} \\ QF_{ij} &= -(V_i^2 - V_i V_j \cos \theta_{ij}) b_{ij} - V_i V_j \sin \theta_{ij} g_{ij} \end{aligned} \quad (2)$$

where $G_{ij}$ and $B_{ij}$ represent the real and imaginary parts of the $i$ th row and $j$ th column of the admittance matrix, respectively.

From a data-driven point of view, PF equations can be built using a data mining technique instead of an admittance matrix. In this work, the identification of the coefficient in the PF equations can be seen as a multi-parameter regression for a high-dimensional surface based on historical operation data. The obtained PF equations from the regression can be further used in the PF calculation, control or operation in the same way as traditional model-based PF equations.

Current studies on the linearization of PF equations show that even though the expression of PF equations is non-linear, it has a high degree of linearity such that the linearized model does not lose too much accuracy. Therefore, we can deduce that the high-dimensional surface described by the ACPF can be approximated as a hyperplane. Though it will introduce errors, the linearity of the model would provide further convenience for the fitting and extension of the model.

### B. PF Linearization Visualization

To visualize the linearization of PF equations in a hyperplane, a simple two-bus system is illustrated, as shown in Fig. 1, where the PF equations can be shown in three-dimensional space.

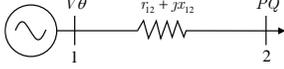

Fig. 1. An illustrative two-bus system

We compare the ACPF with two representative linearized PF models: 1) traditional DCPF equations and 2) the decoupled linear power flow (DLPF) equations proposed in [8]. The formulation of DLPF is shown in (3), where $\mathbf{B'}$ represents the imaginary part of the admittance matrix without shunt elements.

$$\begin{bmatrix} \mathbf{P} \\ \mathbf{Q} \end{bmatrix} = - \begin{bmatrix} \mathbf{B'} & -\mathbf{G} \\ \mathbf{G} & \mathbf{B} \end{bmatrix} \begin{bmatrix} \boldsymbol{\theta} \\ \mathbf{V} \end{bmatrix} \quad (3)$$

In this two-bus system, there are only two PF equations and two independent variables. The independent variables include the active and reactive power injection of bus #2 $P_2$ and $Q_2$. The dependent variables include the voltage magnitude and angle of bus #2 $V_2$ and $\theta_2$. Bus #1 is the reference bus. Fig. 2 shows the value of $\theta_2$ with different combinations of $P_2$ and $Q_2$, calculated by ACPF, DLPF, and DCPF, respectively. Three conclusions can be observed from Fig. 2:

1) The non-linear ACPF surface has a high degree of linearity, which suggests that it can be well described by linear regression.

2) The approximation of DLPF is closer to ACPF than the approximation of DCPF is.

3) The two model-based linear approximations (DCPF and DLPF) still result in clear errors, which suggests that the data-driven linearization approaches still have much to improve on with regards to accuracy.

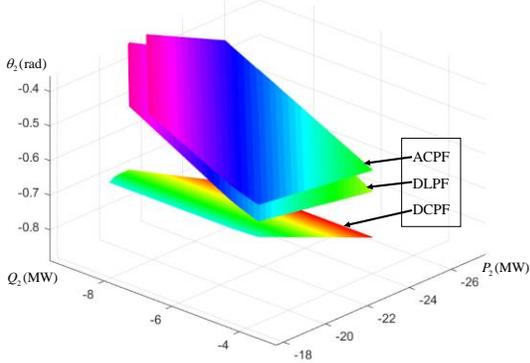

Fig. 2. Visualization of ACPF, DCPF, and DLPF in a two-bus system

### C. PF Mapping Directions

In this paper, the linearization between $P,Q$ and $\theta,V$ is considered. Other relationships (e.g., between $P,Q$, and $V^2$ [7] and $V^2\theta$ [6]) are also available and can be similarly handled.

To explore the linearized mapping rule between $P,Q$, and $\theta,V$, our first attempt is to regress the parameters mapping from $\theta,V$ to $P,Q$, mathematically:

$$P,Q = f(\theta,V) \quad (4)$$

This direction is consistent with the mapping direction of the ACPF equations in (2), where the function of $P,Q$ with respect to $\theta,V$ has an explicit expression. We name this type of mapping direction forward regression.

We also consider the mapping direction from $P,Q$ to $\theta,V$. Such a mapping direction is in accordance with the procedure of the PF calculation, where $P,Q$ are known and $\theta,V$ are to be calculated. We name this mapping direction inverse regression.

### D. Challenges of Regression

The challenges of such a regression lay in two main aspects: to address the collinearity of data and to avoid overfitting. First, collinearity among the voltage angle and magnitude data is inevitable because of the similar rise and fall patterns among the different buses [17]. This will result in ill-conditioned regression and larger errors of PF calculation. Second, the number of variables in the regression parameter matrices for large power systems may be far greater than the amount of historical operation data that represents the current system situation. Such a characteristic may incur the problem of overfitting. Although the performance of an overfitted model may perform surprisingly well on the training dataset, the accuracy may suffer a great loss on the testing dataset [18]. To overcome these two challenges, a PLS-based regression is proposed to ensure the calculation accuracy under collinearity, and a BLR-based regression is proposed to avoid the overfitting of the regressed parameters.

### E. Framework of Data-driven PF Linearization

Based on the above discussions, a framework for data-driven PF linearization is proposed in Fig. 3. The framework is divided into three parts: linearization models, regression methods, and data types.

First, two linearization models, the forward regression model and the inverse regression model, are proposed for different purposes for the PF calculation. Second, both PLS- and BLR-based regression methods are established for both regressions. Finally, different data types, random data under a Monte Carlo simulation and data with collinearity from the public dataset, are tested to illustrate the validity of the proposed models. Each part is detailed in the following sections.

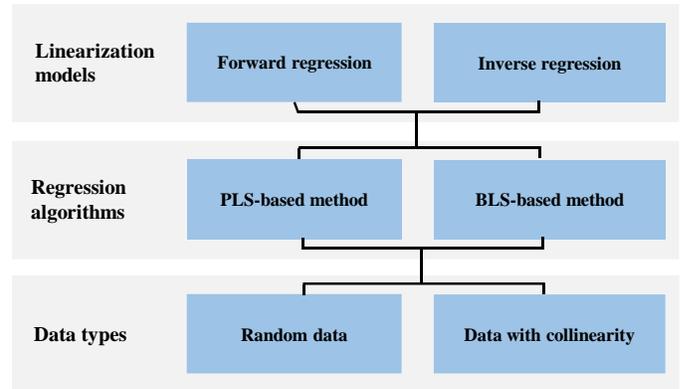

Fig. 3. Framework of data-driven power flow linearization

## III. POWER FLOW LINEARIZATION MODELS

In this section, the models of forward regression and inverse regression are formulated. A theoretical derivation is conducted to reveal the relationship between the regressed matrices and several power system matrices.

### A. Forward Regression Model

The generalized linearization equations of the forward regression model are shown in (5), where $\mathbf{C}_P$ and $\mathbf{C}_Q$ are constant terms of active and reactive bus injections.

$$\begin{bmatrix} \mathbf{P} \\ \mathbf{Q} \end{bmatrix} = \begin{bmatrix} \mathbf{H} & \mathbf{N} \\ \mathbf{M} & \mathbf{L} \end{bmatrix} \begin{bmatrix} \boldsymbol{\theta} \\ \mathbf{V} \end{bmatrix} + \begin{bmatrix} \mathbf{C}_P \\ \mathbf{C}_Q \end{bmatrix} \quad (5)$$

Although the ACPF equations do not have any constant terms, $\mathbf{C}_P$ and $\mathbf{C}_Q$ are added to the linearization equations to enhance the regression capability of the model. In power system operation, the value of some independent variables may remain unchanged, and the regression coefficients of these independent variables in H, N, M, and L may not be regressed. The influences of these independent variables can be absorbed in this constant terms.

The potential application of forward regression is introduced in [12]. It can be used in the PF analysis of a distribution grid more accurately than the traditional model-based PF calculation can because the former considers the invisible control actions of active controllers in the distribution network by learning from historical data.

### B. Inverse Regression Model

It is proposed that inverse regression can calculate PF when considering different bus types, e.g., $PQ, PV, V\theta$ buses. The known and unknown variables in the PF calculation are different among different types of buses. Moreover, the bus types may change from one to another during the calculation process. Our goal is to find the regression model that can obtain the mapping of all of the known variables to the unknown variables for various conditions.

We arrange different types of buses in the following sequence: $PQ, PV, V\theta$.

$$\mathbf{P} = \begin{bmatrix} \mathbf{P}_L^T & \mathbf{P}_S^T & \mathbf{P}_R^T \end{bmatrix}^T \quad \mathbf{Q} = \begin{bmatrix} \mathbf{Q}_L^T & \mathbf{Q}_S^T & \mathbf{Q}_R^T \end{bmatrix}^T$$
$$\mathbf{V} = \begin{bmatrix} \mathbf{V}_L^T & \mathbf{V}_S^T & \mathbf{V}_R^T \end{bmatrix}^T \quad \boldsymbol{\theta} = \begin{bmatrix} \boldsymbol{\theta}_L^T & \boldsymbol{\theta}_S^T & \boldsymbol{\theta}_R^T \end{bmatrix}^T \quad (6)$$

The inverse regression equations can be expressed as a block matrix form in (7), where $\mathbf{C}_1 \sim \mathbf{C}_6$ are constant terms and $\mathbf{A}_{ij}$ is the regression parameter matrix. It should be noted that during the regression stage, both $\begin{bmatrix} \boldsymbol{\theta}_L & \boldsymbol{\theta}_S & \mathbf{P}_R & \mathbf{V}_L & \mathbf{V}_S & \mathbf{V}_R \end{bmatrix}^T$ and $\begin{bmatrix} \mathbf{P}_L & \mathbf{P}_S & \mathbf{Q}_L & \mathbf{Q}_S & \mathbf{Q}_R \end{bmatrix}^T$ are known, and $\mathbf{A}_{ij}$ and $\mathbf{C}_1 \sim \mathbf{C}_6$ are parameters that need to be estimated.

$$\begin{bmatrix} \boldsymbol{\theta}_L \\ \boldsymbol{\theta}_S \\ \mathbf{P}_R \\ \mathbf{V}_L \\ \mathbf{V}_S \\ \mathbf{V}_R \end{bmatrix} = \begin{bmatrix} \mathbf{A}_{11} & \mathbf{A}_{12} & \mathbf{A}_{13} & \mathbf{A}_{14} & \mathbf{A}_{15} \\ \mathbf{A}_{21} & \mathbf{A}_{22} & \mathbf{A}_{23} & \mathbf{A}_{24} & \mathbf{A}_{25} \\ \mathbf{A}_{31} & \mathbf{A}_{32} & \mathbf{A}_{33} & \mathbf{A}_{34} & \mathbf{A}_{35} \\ \mathbf{A}_{41} & \mathbf{A}_{42} & \mathbf{A}_{43} & \mathbf{A}_{44} & \mathbf{A}_{45} \\ \mathbf{A}_{51} & \mathbf{A}_{52} & \mathbf{A}_{53} & \mathbf{A}_{54} & \mathbf{A}_{55} \\ \mathbf{A}_{61} & \mathbf{A}_{62} & \mathbf{A}_{63} & \mathbf{A}_{64} & \mathbf{A}_{65} \end{bmatrix} \begin{bmatrix} \mathbf{P}_L \\ \mathbf{P}_S \\ \mathbf{Q}_L \\ \mathbf{Q}_S \\ \mathbf{Q}_R \end{bmatrix} + \begin{bmatrix} \mathbf{C}_1 \\ \mathbf{C}_2 \\ \mathbf{C}_3 \\ \mathbf{C}_4 \\ \mathbf{C}_5 \\ \mathbf{C}_6 \end{bmatrix} \quad (7)$$

When obtaining all of the parameters in $\mathbf{A}_{ij}$ and $\mathbf{C}_1 \sim \mathbf{C}_6$, the PF calculations can be conducted. When calculating the PF, $\boldsymbol{\theta}_L$, $\boldsymbol{\theta}_S$, $\mathbf{P}_R$ and $\mathbf{V}_L$ in (7) are unknown variables, whereas $\mathbf{V}_S$ and $\mathbf{V}_R$ are known variables. Similarly, as for the independent variables, $\mathbf{P}_L$, $\mathbf{P}_S$ and $\mathbf{Q}_L$ are known, whereas $\mathbf{Q}_S$ and $\mathbf{Q}_R$ are unknown. Hence, the equation in (7) can be rewritten in the form of (8), where $\mathbf{x}_1 = [\mathbf{P}_L, \mathbf{P}_S, \mathbf{Q}_L]^T$ and $\mathbf{y}_2 = [\mathbf{V}_S, \mathbf{V}_R]^T$ are known variables, and $\mathbf{x}_2 = [\mathbf{Q}_S, \mathbf{Q}_R]^T$ and $\mathbf{y}_1 = [\boldsymbol{\theta}_L, \boldsymbol{\theta}_S, \mathbf{P}_R, \mathbf{V}_L]^T$ are unknown variables. After obtaining all of the parameters from the regression, the unknown variables can be calculated in (9). The invertibility of matrix $\tilde{\mathbb{A}}_{22}$ is discussed in later sections.

$$\begin{bmatrix} \mathbf{y}_1 \\ \mathbf{y}_2 \end{bmatrix} = \begin{bmatrix} \tilde{\mathbb{A}}_{11} & \tilde{\mathbb{A}}_{12} \\ \tilde{\mathbb{A}}_{21} & \tilde{\mathbb{A}}_{22} \end{bmatrix} \begin{bmatrix} \mathbf{x}_1 \\ \mathbf{x}_2 \end{bmatrix} + \begin{bmatrix} \tilde{\mathbb{C}}_1 \\ \tilde{\mathbb{C}}_2 \end{bmatrix} \quad (8)$$

$$\mathbf{x}_2 = \tilde{\mathbb{A}}_{22}^{-1} (\mathbf{y}_2 - \tilde{\mathbb{A}}_{21} \mathbf{x}_1 - \tilde{\mathbb{C}}_2)$$
$$\mathbf{y}_1 = \tilde{\mathbb{A}}_{11} \mathbf{x}_1 + \tilde{\mathbb{A}}_{12} \mathbf{x}_2 + \tilde{\mathbb{C}}_1 \quad (9)$$

The reasons for building the regression equation in (7) in such form are twofold:

1) To maintain the feasibility of the PF calculation with flexible bus type settings.

The model of inverse regression is applicable for different bus type settings because all buses are reordered and calculated in the same sequence of $PQ, PV, V\theta$. When the bus types transform from one into another, the regression parameter matrix $\mathbf{A}_{ij}$ in (7) is reordered rather than recalculated.

The necessary condition of such ability comes from the fact that $\tilde{\mathbb{A}}_{22}$ in (8) is reversible. In the inverse regression model, the independent variables corresponding to $\tilde{\mathbb{A}}_{22}$ are reactive power injections of the $PV$ and $V\theta$ buses. The dependent variables corresponding to $\tilde{\mathbb{A}}_{22}$ are the voltage magnitude of the $PV$ and $V\theta$ buses. These quantities are not constants in the historical data. Regarding the voltage magnitude, the fluctuation of $PV$ and $V\theta$ buses is inevitable. The maximum fluctuation range of each bus depends on the voltage control device (e.g., the maximum range is set as 0.05p.u.-0.215p.u. in continental Europe [19]). Therefore, the obtained $\tilde{\mathbb{A}}_{22}$ is a full ranked matrix. In contrast, matrices $\mathbf{H}, \mathbf{N}, \mathbf{M}, \mathbf{L}$ in (5) obtained from forward regression cannot be used to obtain the mapping via the formulation of (7)-(9). This is because the $\tilde{\mathbb{A}}_{22}$ in the forward regression corresponds to the dependent variables of the $PQ$ and $PV$ buses. When there are zero-power-injected $PQ$ buses, the regression cannot obtain a full ranked or nonsingular $\tilde{\mathbb{A}}_{22}$, so the mapping can hardly be obtained.

2) To remove $\mathbf{P}_R$ from independent variables to avoid collinearity.

In (7), all the active and reactive power injections are independent variables, except the active power injection of the $V\theta$ bus $\mathbf{P}_R$, because the relationship of active injection of all buses can be approximated in (10).

$$\sum_i P_i \approx 0 \quad (10)$$

In other words, $\mathbf{P}_R$ can be almost determined by active power injections of other nodes. The regression will lead to collinearity and will result in the ill-conditioned regression parameter matrix $\mathbf{A}_{ij}$. Although the relation in (10) is not a strict equation, it results in the problem of collinearity. For more on the impact of collinearity on the regression, refer to [20, 21]. Instead, the regression model for $\mathbf{P}_R$ is added as the third row in (7). Such a formulation can consider the power balance of the power system and indirectly accounts for network losses by introducing the item of reactive injections and constant terms into the independent variables.

*C. Relationship with Physical Parameter Matrices*

Interestingly, the value of forward and inverse regression parameter matrices is numerically similar to the value of several power system matrices. The derivation process of this relationship is shown in Fig. 4.

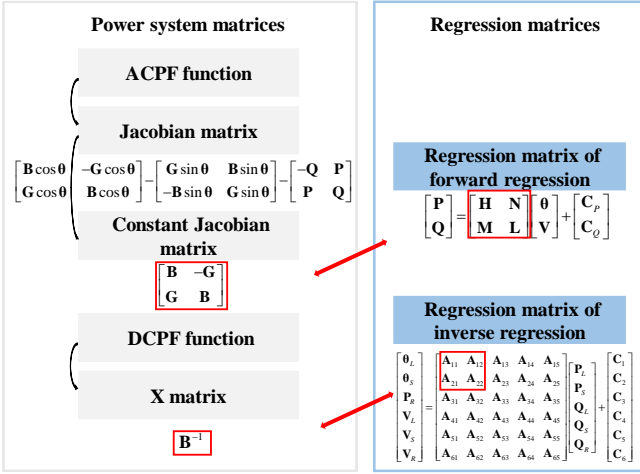

Fig. 4. Derivation process of relationships between regression parameter matrices and several power system matrices

Fig. 4 presents a theoretical derivation of how the regression parameter matrices are related to several power system matrices. First, the forward regression can be seen as the first-order Taylor's approximation of the PF equations in (2). Therefore, the forward regression parameter matrix can be seen as the partial derivative of the PF equations: the Jacobian matrix. However, given a set of power system operation data, the value of the Jacobian matrix is different for different operating points, whereas the value of the regression parameter matrix is constant. Hence, the constant Jacobian matrix that is widely used in the Newton-Raphson method is a reasonable approximation of the forward regression parameter matrix.

Second, it is complicated to derive the theoretical explanation of the inverse regression parameter matrix from the partial derivative of the PF equations because $\mathbf{\theta}$, $\mathbf{V}$ are difficult to represent as a function of $\mathbf{P}$, $\mathbf{Q}$ with definite formulations. Thus, the derivation of the inverse partial derivative (e.g., $\partial\mathbf{\theta}/\partial\mathbf{P}$, $\partial\mathbf{\theta}/\partial\mathbf{Q}$, $\partial\mathbf{V}/\partial\mathbf{P}$, $\partial\mathbf{V}/\partial\mathbf{Q}$) from the PF equations require implicit differentiation. From the four inverse partial derivatives, $\partial\mathbf{\theta}/\partial\mathbf{P}$ can be easily approximated by the inverse matrix of $\mathbf{B}$ according to the DCPF equations:

$$\mathbf{P}=\mathbf{B}\mathbf{\theta} \qquad (11)$$

The approximation of $\partial\mathbf{\theta}/\partial\mathbf{P}$ corresponds to the matrices $\mathbf{A}_{11}$, $\mathbf{A}_{12}$, $\mathbf{A}_{21}$ and $\mathbf{A}_{22}$ in (7). These relationships discussed above can serve as an indicator of overfitting.

*D. Mapping of Branch Power Flow*

The mapping of the branch PF is similar to the mapping of the power injection. Given the historical data of active and reactive branch PF ($PF$ and $QF$), the mapping rule can be regressed. The mapping direction can either be from $P,Q$ to $PF,QF$ or from $\theta,V$ to $PF,QF$. Taking the former as an example, the linearization equations is in (12), where $\mathbf{P}_R$ is removed from the independent variables:

$$\begin{bmatrix}\mathbf{PF}\\ \mathbf{QF}\end{bmatrix} = \begin{bmatrix}\mathbf{H}_{line} & \mathbf{N}_{line}\\ \mathbf{M}_{line} & \mathbf{L}_{line}\end{bmatrix}\begin{bmatrix}\mathbf{P}\\ \mathbf{Q}\end{bmatrix} + \begin{bmatrix}\mathbf{C}_{PF}\\ \mathbf{C}_{QF}\end{bmatrix} \qquad (12)$$

## IV. REGRESSION ALGORITHMS

The mathematical models of the forward regression, inverse regression, and branch PF regression are linear regression models. To simplify the representation, the generalized regression equation is presented by $\mathbf{A}$, $\mathbf{X}$ and $\mathbf{Y}$, which represent the regression parameter matrix, matrix of independent variables and matrix of dependent variables, respectively.

$$\mathbf{Y} = \mathbf{A}\mathbf{X} \qquad (13)$$

In the data-driven PF regression, $\mathbf{X}$ and $\mathbf{Y}$ are matrices rather than vectors. The number of columns represents different datasets at different times (or different operation snapshots in a power system), and the number of rows represents different variables. Taking forward regression as an example, $\mathbf{X}$ and $\mathbf{Y}$ are formulated in (14), where the last row of $\mathbf{X}$ corresponds to the constant terms.

$$\mathbf{X} = \begin{bmatrix}\mathbf{\theta}^1 ... \mathbf{\theta}^t ... \mathbf{\theta}^T\\ \mathbf{V}^1 ... \mathbf{V}^t ... \mathbf{V}^T\\ 1 ... 1 ... 1\end{bmatrix}, \mathbf{Y} = \begin{bmatrix}\mathbf{P}^1 ... \mathbf{P}^t ... \mathbf{P}^T\\ \mathbf{Q}^1 ... \mathbf{Q}^t ... \mathbf{Q}^T\end{bmatrix} \qquad (14)$$

In the process of theoretical derivation, $\mathbf{A}$, $\mathbf{X}$, and $\mathbf{Y}$ are expressed in the form of rows:

$$\mathbf{X} = \begin{bmatrix}\mathbf{x}_1 & \mathbf{x}_2 & ... & \mathbf{x}_N\end{bmatrix}^T \mathbf{Y} = \begin{bmatrix}\mathbf{y}_1 & \mathbf{y}_2 & ... & \mathbf{y}_M\end{bmatrix}^T \mathbf{A} = \begin{bmatrix}\mathbf{a}_1 & \mathbf{a}_2 & ... & \mathbf{a}_M\end{bmatrix}^T \quad (15)$$

In the following sub-section, both PLS-based and BLR-based algorithms are proposed.

*A. PLS-Based Algorithm*

The objective of the PLS-based algorithm is to regress between two zero-mean data blocks, $N \times T$ matrix $\mathbf{X}$ and $M \times T$ matrix $\mathbf{Y}$. It can address the collinearity and lack of observations after combining the features from principal component analysis (PCA) and canonical correlation analysis (CCA) [22]. PLS decomposes $p$ components from the matrix of independent variables $\mathbf{X}$ and matrix of dependent variables $\mathbf{Y}$ into the form

$$\mathbf{X} = \mathbf{C}\mathbf{T}^T + \mathbf{E}$$
$$\mathbf{Y} = \mathbf{R}\mathbf{U}^T + \mathbf{F} \qquad (16)$$

where $\mathbf{T},\mathbf{U}$ are $T \times p$ matrices of the $p$ extracted components, $\mathbf{C},\mathbf{R}$ are $N \times p$ and $M \times p$ matrices represent the loading matrices, and $\mathbf{E}, \mathbf{F}$ are $N \times T$ and $M \times T$ matrices and represent the residuals. The datasets of $\mathbf{X}$ and $\mathbf{Y}$ share a similar rise and fall pattern in different rows, which corresponds to the collinearity in the power system data (e.g., active power injections tend to rise and fall at the same time in different buses). In (16), PLS projects $\mathbf{X}$ and $\mathbf{Y}$ onto two small matrices $\mathbf{T}$ and $\mathbf{U}$ to extract the key components that $\mathbf{Y}$ correlate to $\mathbf{X}$.

Calculation of the correlated matrices is based on the nonlinear iterative partial least squares algorithm [23]. Finally, given the matrices of $\mathbf{X}^*$ and $\mathbf{Y}^*$ as the updated independent and dependent variables, the matrix of dependent variables is predicted in the form

$$\mathbf{Y}^* = \mathbf{A}\mathbf{X}^* \text{ where } \mathbf{A}^T = \mathbf{X}^T \mathbf{U}(\mathbf{T}^T \mathbf{X}\mathbf{X}^T \mathbf{U})^{-1} \mathbf{T}^T \mathbf{Y} \quad (17)$$

*B. BLR-based Algorithm*

The BLR-based algorithm is conducted within the context of Bayesian inference [24]. Different vectors of $\mathbf{Y}$ in (15) are regressed. Taking $\mathbf{y}_i$ as an example, (18) represents the regression equation

$$\mathbf{y}_i = \mathbf{a}_i \mathbf{X} + \mathbf{e}_i, \ i = 1, 2, ..., M \quad (18)$$

where $\mathbf{e}_i$ represents the additive noise of $\mathbf{y}_i$, and $\mathbf{X}$, $\mathbf{y}_i$ are centered in a previous step. Each $\mathbf{a}_i$ represents a vector:

$$\mathbf{a}_i = \begin{bmatrix} a_{i1} & ... & a_{ij} & ... & a_{iL} \end{bmatrix} \quad (19)$$

According to the Bayesian inference framework, the posterior probability of $\mathbf{a}_i$ follows

$$p(\mathbf{a}_i | \mathbf{y}_i, \mathbf{X}) \propto p(\mathbf{a}_i) p(\mathbf{y}_i, \mathbf{X} | \mathbf{a}_i) \quad (20)$$

where $p(\mathbf{a}_i)$ represents the prior and $p(\mathbf{y}_i, \mathbf{X} | \mathbf{a}_i)$ represents the likelihood. The prior is introduced to avoid overfitting by setting a simple form of presumption on the posterior distribution. In this work, an elliptical Gaussian distribution prior for $\mathbf{a}_i$ is assumed:

$$p(\mathbf{a}_i) = \prod_{j=0}^{L} N(a_{ij} | 0, \beta_j^{-1}) \quad (21)$$

Each $a_j$ has its own standard deviation $\beta_j^{-1}$ that can adjust according to the real observation data of $a_j$. The distribution of the reciprocal of standard deviations $\boldsymbol{\beta}$ can be proven to follow a Gamma distribution, which can be described by two hyperparameters: $\lambda_1$ and $\lambda_2$. The benefit of giving the hyperparameter prior is that the hierarchical formulation encourages sparsity over the prior than that of a flat hierarchy Gaussian prior [25]. With the assumption of a Gaussian distribution with additive noise $\mathbf{e}_i$, the likelihood can be written as

$$p(\mathbf{y}_i, \mathbf{X} | \mathbf{a}_i) = (2\pi \beta_j^{-2})^{-L/2} \exp\{-\frac{\beta_j^2}{2} \|\mathbf{y}_i - \mathbf{X}\mathbf{a}_i\|^2\} \quad (22)$$

To calculate parameter $\mathbf{a}_i$, a maximum a posterior (MAP) optimization is conducted. During the iteration of the optimization process, the estimation of $a_j$ is set to zero when its deviation $\beta_j^{-1}$ is under a certain threshold. This gives a flexibility of adjustment on the sparsity. The reasonable sparsity is essential for forward regression. Because the regression parameter matrix shows similar patterns with the constant Jacobian matrix, which is a sparse matrix. The detailed derivation, optimization process and proving of convergence are in reference [25].

V. EXPERIMENTAL RESULTS

*A. Data Generation*

Data of the power system operation measurement used in the case studies are divided into two categories: Monte Carlo simulation and public testing data. Data processing from both categories is performed in MATLAB with the aid of MATPOWER 6.0 [26]. Parameters are regressed using the training dataset, and the PF calculation accuracy is tested using the newly generated testing dataset.

1) Monte Carlo simulation

A Monte Carlo simulation was run on meshed transmission grids, which include IEEE 5, 30, 57, and 118-bus systems and radial distribution grids, which include the IEEE 33-bus system [27] and the modified 123-bus system [28]. The active load consumption is calculated from the preset load consumption multiplied by a factor randomly drawn from a uniform distribution over the interval [0.8, 1.2]. The reactive load consumption is calculated from the active load consumption multiplied by a factor randomly drawn from a uniform distribution over the interval [0.15, 0.25].

2) Public testing data

We use the hourly load data of the NREL-118 test system [29] to replace the randomly generated active load consumption in the Monte Carlo simulation. The load data are synthetic that have similar rise and fall patterns that were learned from 1980-2012 weather and load data. Gaussian noise is added to the original load data because the original load data is divided on a pro-rata basis at different times. A scale factor is multiplied by the load data to balance the system generation capacity. The data of the NREL-118 test system are used to test the model adaptability to the data with collinearity.

*B. Basic Results*

We first fit the data-driven PF equations on the training dataset using the proposed regression algorithms. Then, the PF calculation is conducted on the testing dataset. The average errors of the proposed PLS- and BLR-based algorithms compared with both the DCPF and DLPF methods on forward calculation, inverse calculation, and branch PF calculation are shown in TABLE I.

The accuracy of forward regression is measured by the errors of power injection $P$ and $Q$ in all buses; the accuracy of inverse regression is measured by the errors of voltage magnitude $V$ in $PQ$ buses and voltage angle $\theta$ in all buses except the $V\theta$ bus. As is illustrated in TABLE I, several conclusions can be drawn:

1) Among all cases of PF calculations, the proposed data-driven approaches are more accurate than or at least as accurate as that of model-based DCPF and DLPF methods.

TABLE I. ERRORS OF FORWARD, INVERSE AND BRANCH CALCULATION ON DIFFERENT CASES

| Cases | Size of training data | Size of testing data | Forward calculation | | | | | Inverse calculation | | | | Branch PF calculation | | | | |
|---|---|---|---|---|---|---|---|---|---|---|---|---|---|---|---|---|
| | | | Errors | DCPF | DLPF | PLS | BLR | Errors | DLPF | PLS | BLR | Errors | DCPF | DLPF | PLS | BLR |
| IEEE 5 | 100 | 300 | P | 24.11 | 1.117 | **0.412** | 0.615 | θ | 0.020 | **8.2e-4** | 6.8e-3 | PF | 8.120 | 8.120 | **0.126** | 0.609 |
| | | | Q | --- | 66.21 | **0.940** | 1.065 | V | 7.8e-4 | **2.0e-5** | 4.1e-3 | QF | --- | --- | **8.934** | 256.1 |
| IEEE 30 | 100 | 300 | P | 12.49 | 0.578 | **0.034** | 0.238 | θ | 0.154 | **1.9e-3** | 0.071 | PF | 7.734 | 7.562 | **0.104** | 0.825 |
| | | | Q | --- | 12.66 | **0.404** | 0.471 | V | 9.9e-4 | **1.0e-5** | 1.4e-3 | QF | --- | --- | **1.340** | 226.9 |
| IEEE 33 | 100 | 300 | P | 67.05 | 1.114 | **0.012** | 0.012 | θ | 0.028 | **4.3e-4** | 0.011 | PF | 1.142 | 1.142 | **5.0e-3** | 8.8e-3 |
| | | | Q | --- | 0.759 | 0.044 | **0.027** | V | 2.0e-3 | **7.3e-6** | 6.5e-4 | QF | --- | --- | **0.013** | 0.497 |
| IEEE 57 | 300 | 300 | P | 98.11 | 7.343 | **0.262** | 2.132 | θ | 0.215 | **0.036** | 0.218 | PF | 19.16 | 13.22 | **0.395** | 0.965 |
| | | | Q | --- | 26.83 | **0.300** | 2.990 | V | 7.1e-3 | **2.1e-4** | 1.1e-3 | QF | --- | --- | **5.227** | 24.71 |
| IEEE 118 | 300 | 300 | P | 16.89 | 4.546 | **0.061** | 1.385 | θ | 2.593 | **0.074** | 0.296 | PF | 86.96 | 86.04 | **2.263** | 7.078 |
| | | | Q | --- | 77.85 | **1.096** | 31.73 | V | 1.9e-3 | **1.2e-4** | 8.1e-4 | QF | --- | --- | **5.570** | 68.27 |
| NREL 118 | 300 | 300 | P | 85.90 | 9.486 | **0.161** | 1.207 | θ | 3.003 | 0.622 | **0.271** | PF | 33.08 | 29.37 | 10.59 | **4.326** |
| | | | Q | --- | 107.4 | **0.486** | 3.982 | V | 2.3e-3 | **6.3e-4** | 7.6e-4 | QF | --- | --- | **28.07** | 36.53 |
| Modified 123 | 300 | 300 | P | 12.49 | 0.512 | **0.007** | 0.452 | θ | 0.091 | **3.2e-4** | 2.6e-3 | PF | 0.319 | 0.319 | **5.1e-4** | 6.6e-3 |
| | | | Q | --- | 2.071 | **0.003** | 0.003 | V | 2.3e-3 | **3.2e-6** | 3.5e-6 | QF | --- | --- | **3.6e-3** | 7.4e-3 |

✧ The errors of P, Q, PF, and QF are in mean absolute percentage error with the unit of 100%, whereas the errors of θ and V are in mean absolute error.
✧ The errors of Q correspond to DCPF are not shown because DCPF is not able to calculate reactive power. The errors of QF that correspond to DCPF and DLPF are not shown because DCPF and DLPF cannot calculate the reactive power flow.

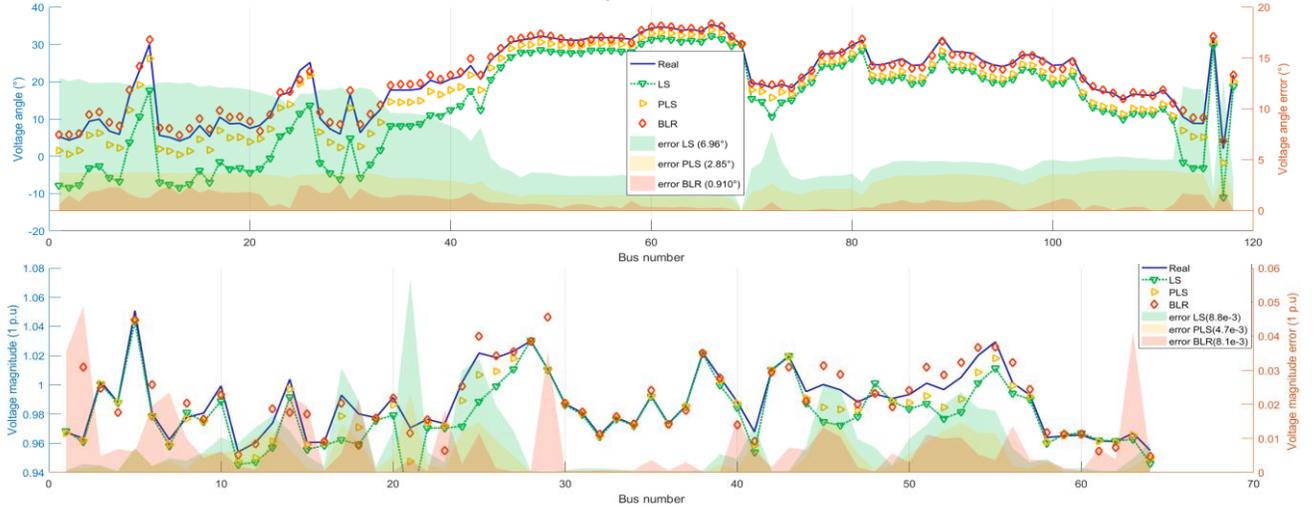

Fig. 5. Voltage angles and magnitudes of NREL 118 test system

2) The proposed data-driven approaches provide better results in areas where model-based methods are not as accurate, such as reactive power injection $Q$ of forward calculations and the active power flow $PF$. The errors in these areas are more than one order less than that of model-based methods.

3) In most of the cases (IEEE 5, 30, 57, 118, modified 123-bus systems), the PLS-based algorithm is more accurate than the BLR-based algorithm. The BLR-based algorithm only demonstrates better results on IEEE 33-bus in $P,Q$ calculations and NREL-118 test systems in $\theta, PF$ calculations.

### C. Calculation Results under Data Collinearity

The inverse calculation results of the NREL-118 test system are illustrated in Fig. 5 to present a visualization of the computation accuracy. The calculation error presented in Fig. 5 is a group of results from 300 groups of testing data. To show the robustness of the algorithm, the error in Fig.5 is the largest among all 300 groups in the NREL-118 test system. Only the $PQ$ buses are shown for the voltage magnitude results. To test the algorithm efficiency under data collinearity, the least square (LS) regression does not consider data collinearity is applied as a contrast. It is clear that the proposed algorithm performs better than the LS algorithm, particularly in the voltage angle calculation. The PLS-based algorithm is more accurate than the BLR-based algorithm in terms of voltage magnitude, whereas the BLR-based algorithm is more accurate than the PLS-based algorithm in terms of the voltage angle. The results demonstrate the efficiency of the proposed regression algorithms under data collinearity.

### D. Regression Parameters

To verify the relationship between regression parameter matrices and several power system matrices, IEEE 5 and 57-bus systems are analyzed. The forward regression parameter matrix of the IEEE 5-bus system based on the PLS and BLR-based algorithms compared with the constant Jacobian matrix are shown in Fig. A1 (see appendix). Regarding the inverse regression parameter matrix, $\mathbf{A}_{11}$, $\mathbf{A}_{12}$, $\mathbf{A}_{21}$ and $\mathbf{A}_{22}$ in (7) compared with the inverse matrix of $\mathbf{B}$ in DCPF are also shown in Fig. A1. Similarly, the same comparisons of the IEEE 57-bus system are shown in Fig. A2 (see appendix). Regarding the IEEE 5-bus system, the forward regression parameter matrices of both the PLS- and BLR-based algorithms are extremely similar to the constant Jacobian matrix. The $\mathbf{A}_{11}$, $\mathbf{A}_{12}$, $\mathbf{A}_{21}$ and $\mathbf{A}_{22}$ in the inverse regression of both the PLS and BLR-based algorithms are similar to the inverse matrix of $\mathbf{B}$ in DCPF. These results validate the theoretical analysis in Fig. 4. Regarding the IEEE 57-bus system, the constant Jacobian matrix is highly sparse with diagonal non-zero parameters. The forward regression parameter matrix of the PLS-based algorithm contains the diagonal non-zero parameters and many other off-diagonal large value parameters. This parameter overfitting can be eliminated in the BLS-based algorithm. The inverse matrix of $\mathbf{B}$ in DCPF is a full matrix; thus, the regression parameter matrices of both PLS and BLR-based algorithms provide an acceptable approximation. There are several zero columns because of the zero-power-injected buses. The columns that correspond to these buses are regressed to zero and have no influence on the calculation accuracy as long as the injections of these buses remain zero.

## VI. Conclusion

In this paper, we provide a data-driven PF linearization approach that bridges the gap between model-based PF linearization methods and data-driven power system analysis approaches. Forward and inverse regression methods as well as branch PF mapping are proposed to facilitate a variety of linearized PF calculation. To conquer the collinearity of the data, PLS- and BLR-based regression methods are used. Several cases, including meshed transmission grids, radial distribution grids, and public testing system, are examined. The results verify the distinct advantage on the accuracy of the proposed data-driven approaches over several selected methods. More importantly, the parameter matrices obtained from the regression are found to maintain physical significance of the model-based parameters, which demonstrates its ability to identify the physical reality of the power system.

We envision that the proposed data-driven linearization approach serves as the foundation of accurate linearization calculations and optimization methods.

# Appendix-Data-Driven Power Flow Linearization: A Regression Approach

Yuxiao Liu, *Student Member, IEEE,* Ning Zhang, *Member, IEEE,* Yi Wang, *Student Member, IEEE,* Jingwei Yang, *Student Member, IEEE,* and Chongqing Kang, *Fellow, IEEE*

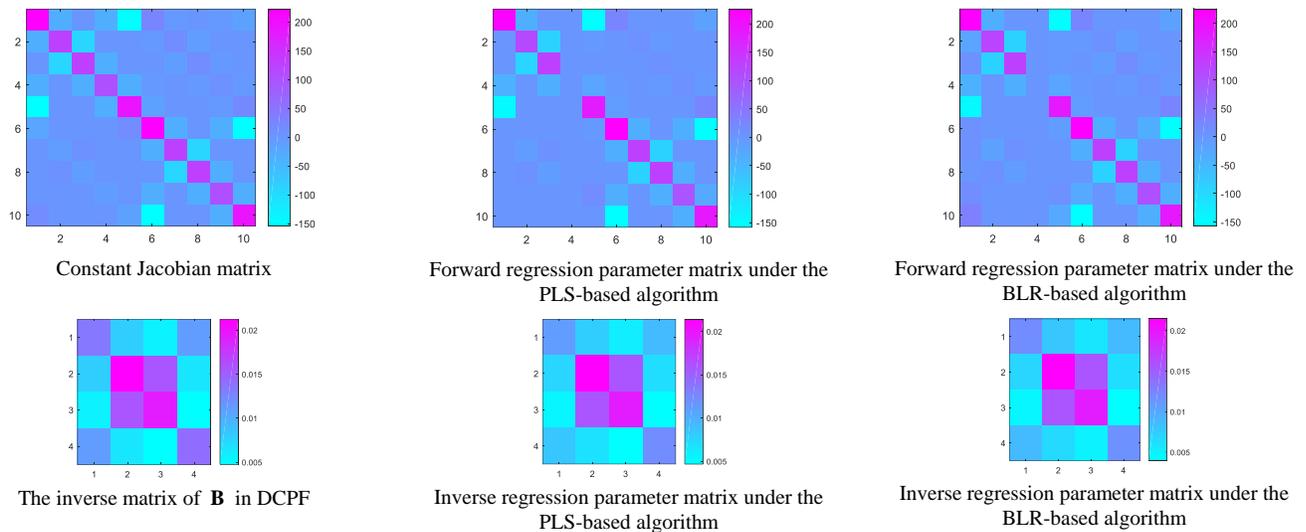

Fig. A1. Comparisons between regression parameter matrices and several power system matrices of IEEE 5-bus system

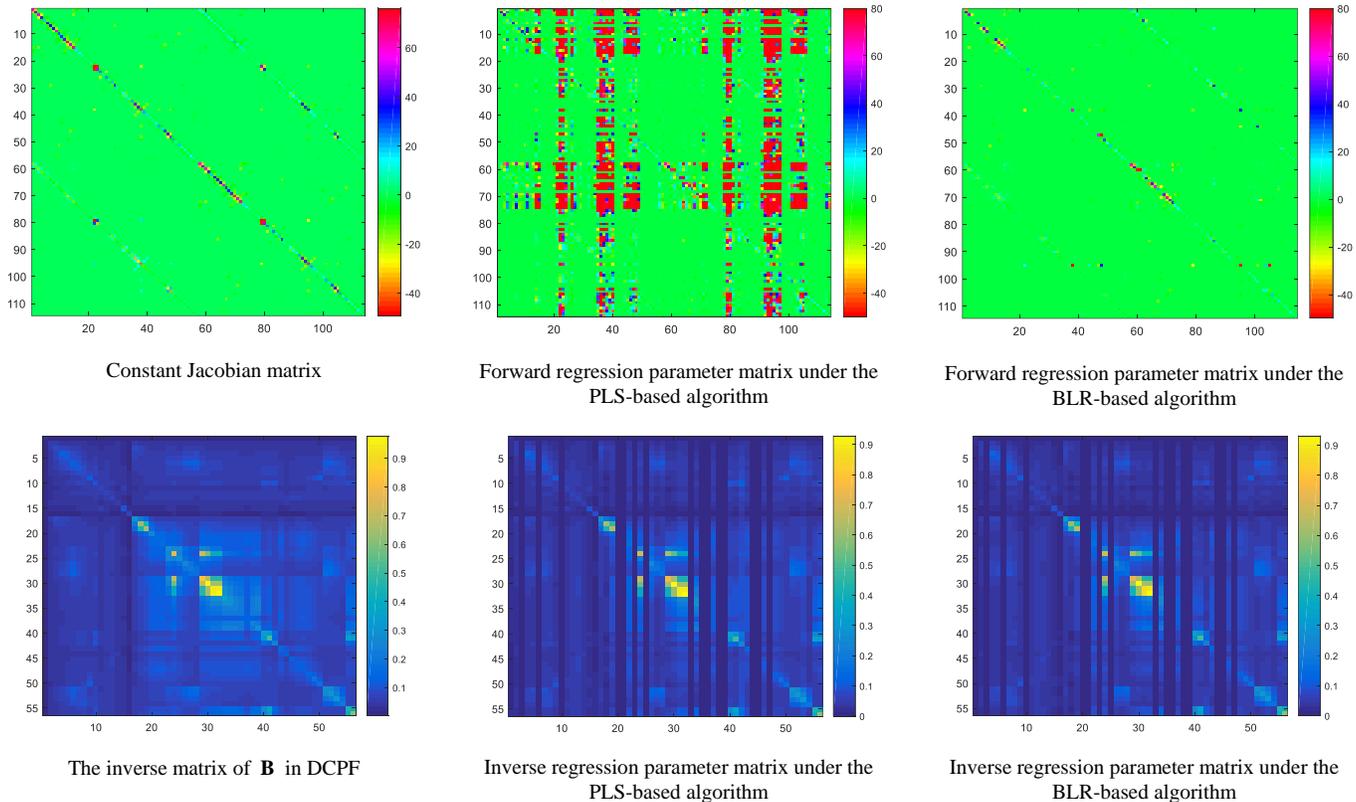

Fig. A2. Comparisons between regression parameter matrices and several power system matrices of IEEE 57-bus system